%% file: aaai2026.tex
\documentclass[letterpaper]{article} 
\usepackage{authblk}
\usepackage{times}  
\usepackage{helvet}  
\usepackage{courier}  
\usepackage[hyphens]{url}  
\usepackage{graphicx} 
\usepackage{natbib}  
\usepackage{caption} 
\usepackage{algorithm}
\usepackage{xcolor}  
\usepackage{multirow}
\usepackage{newfloat}
\usepackage{listings}
\usepackage{makecell} 
\usepackage{booktabs} 
\usepackage{subcaption}  
\usepackage{amsmath} 
\usepackage{amssymb}
\usepackage{amsthm}
\usepackage{algorithmicx}
\usepackage{algpseudocode}
\usepackage{float}

\algnewcommand\algorithmicinput{\textbf{Input:}}
\algnewcommand\Input{\item[\algorithmicinput]}

\algnewcommand\algorithmicoutput{\textbf{Output:}}
\algnewcommand\Output{\item[\algorithmicoutput]}

\newtheorem{proposition}{Proposition}
\usepackage{newfloat}
\usepackage{listings}
\usepackage{newfloat}
\usepackage{listings}
\DeclareCaptionStyle{ruled}{labelfont=normalfont,labelsep=colon,strut=off} 
\floatstyle{ruled}
\newfloat{listing}{tb}{lst}{}
\floatname{listing}{Listing}
\author[1]{ George Panagopoulos}
\author[2]{Johannes F. Lutzeyer}
\author[3]{Sofiane Ennadir}
\author[2]{Michalis Vazirgiannis}
\author[1]{Jun Pang}
\affil[1]{University of Luxembourg, 6 avenue de la Fonte, Esch-sur-Alzette, Luxembourg }
\affil[2]{LIX, École Polytechnique, IP Paris, France}
\affil[3]{KTH Royal Institute Of Technology, SE-100 44 Stockholm, Sweden }
\affil[ ]{\{georgios.panagopoulos, jun.pang\}@uni.lu,johannes.lutzeyer@polytechnique.edu,ennadir@kth.se}
\title{Efficient Data Selection for Training Genomic Perturbation Models}

\usepackage{bibentry}

\begin{document}

\maketitle

\begin{abstract}
Genomic studies face a vast hypothesis space, while interventions such as gene perturbations remain costly and time-consuming. To accelerate such experiments, gene perturbation models predict the transcriptional outcome of interventions. Since constructing the training set is challenging, active learning is often employed in a “lab-in-the-loop” process. While this strategy makes training more targeted, it is substantially slower, as it fails to exploit the inherent parallelizability of Perturb-seq experiments. Here, we focus on graph neural network–based gene perturbation models and propose a subset selection method that, unlike active learning, selects the training perturbations in one shot. Our method chooses the interventions that maximize the propagation of the supervision signal to the model. The selection criterion is defined over the input knowledge graph and is optimized with submodular maximization, ensuring a near-optimal guarantee. Experimental results across multiple datasets show that, in addition to providing months of acceleration compared to active learning, the method improves the stability of perturbation choices while maintaining competitive predictive accuracy.
\end{abstract}


\input{introduction}

\input{related}

\input{methodology}

\input{experiments}
\vspace{-.5cm}
\section{Conclusion}
Genomic experiments cannot exhaust the available intervention options, thus, gene perturbation models are used to predict the outcomes of these interventions. However, their training sets consist of costly genomic experiments themselves, calling for efficient training strategies. In this work, we focused on graph neural networks and propose \textit{\textsc{GraphReach}}, a subset-selection method that chooses perturbations such that the model is trained more efficiently. In the experiments, we observed that compared to existing active learning solutions, \textit{\textsc{GraphReach}} exhibits significant real-world speedup (from five months to one) as well as more stable selections without sacrificing accuracy in general and while being robust to errors in the knowledge graph. We deem \textsc{GraphReach} a valid alternative to random or active-learning-based training given its simple deployment and better overall performance. In the future, we plan to examine hybrid techniques that utilize subset selection methods to jump-start the training up to a point and blend in active learning for the rest of the runs. Moreover, we plan to research the use of subset-selection for non-graph-based methods, covering both foundation models \citep{theodoris2023transfer} and regression \citep{ahlmann2025deep}.

\bibliographystyle{plain}   
\bibliography{aaai26}  

\input{appendix}

\end{document}

%% file: introduction.tex
\section{Introduction}
\label{sec:intro}
Genomic research enables the study of genetic factors underlying various health conditions, opening new avenues for therapeutic development. 
Techniques such as CRISPR interference and PerturbSeq \citep{barrangou2016applications} have revolutionized the landscape of genomic experimentation by enabling high-throughput screenings. 
However, the majority of CRISPR-based PerturbSeq experiments are restricted to hundreds of single gene perturbations \citep{dixit2016perturb,adamson2016multiplexed,peidli2022scperturb} due to budget and time constraints, despite having over 20,000 potential gene targets. To assist exploring this vast space, machine learning models have been developed to predict the outcome of gene perturbations on a given cell \citep{bereket2024modelling,gaudelet2024season,lopez2023learning,lotfollahi2023predicting}. 

\begin{figure*}[h!]
    \centering
    \begin{subfigure}{0.59\textwidth}  
        \centering
    \centering
    \includegraphics[width=0.95\linewidth]{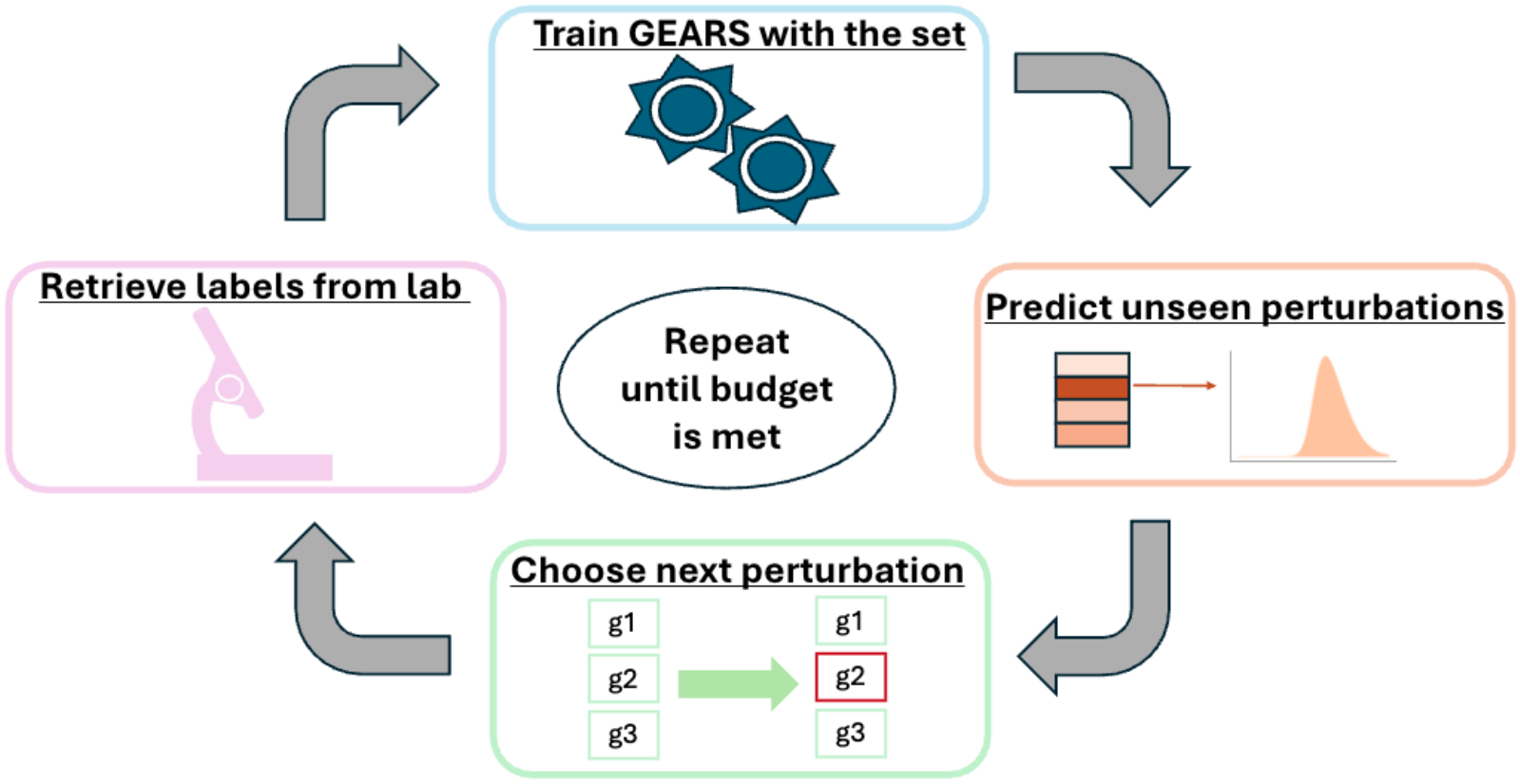}
    \caption{ Active learning.}
    \label{fig:al}
\end{subfigure}
\begin{subfigure}{0.4\textwidth} 
    \centering
    \includegraphics[width=0.45\linewidth]{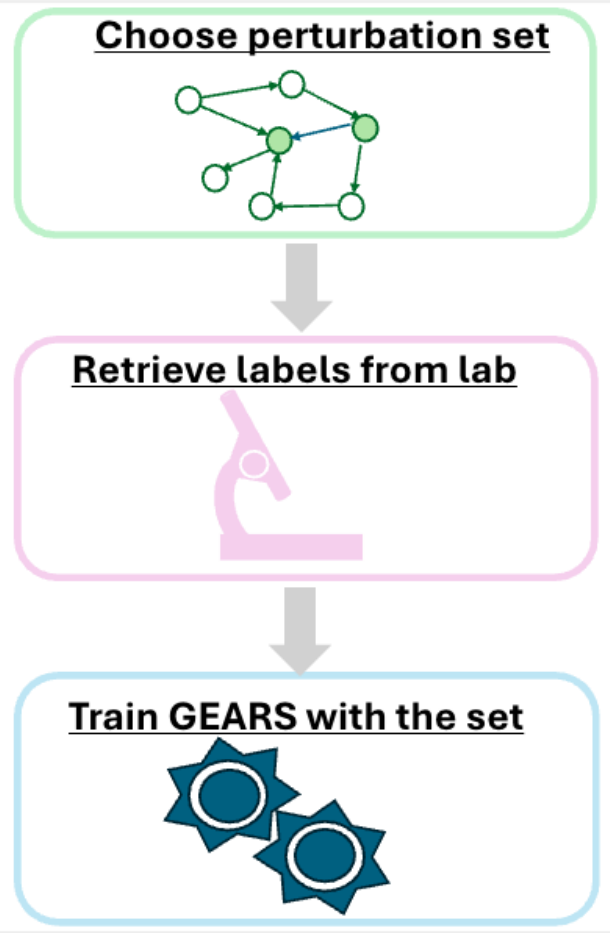}
    \caption{Subset selection.}
    \label{fig:df}
    \end{subfigure}
    \caption{The difference between active learning and subset selection for training GEARS.}
    \label{fig:al_vs_selection}
\end{figure*}

Recent methods often leverage graph-based models that use gene-gene interaction networks for perturbation prediction \citep{roohani2024predicting}. They require a substantial number of training samples that stem from genomic experiments pertaining to a specific hypothesis, rendering training costly and time-consuming because it requires feedback from a wet-lab. 
Active learning has been proposed to address this problem \citep{huang2024sequential} by defining an iterative interaction between the wet lab and the base model to run experiments for genes that will optimize the training, as shown in Fig.~\ref{fig:al}. This ensures that resources are not wasted in non-informative genomic experiments. However, in this setting, the training can take months, because despite the process having typically a few iterations ($\leq 5$) \citep{huang2024sequential}, each iteration translates to a CRISPRi-based PerturbSeq experiment that may take 3-5 weeks \citep{gasperini2019genome,huang2024sequential}, without accounting for operational delays due to communication or computational hurdles like model retraining.

Another drawback of active learning is that it relies on a randomly initialized model to decide the first batch of genes. As a result, different runs can yield substantially different training sets, which undermines reproducibility and limits the reusability of collected data in downstream analyses. To mitigate this instability, we need selection strategies that decouple the model’s output from the training set choice. Density-based active learning is model-independent \citep{hacohen2022active}, but it does not apply here because genes lack predefined features; their embeddings are learned by the model itself, so active learning still operates in a model-dependent space \citep{huang2024sequential}.

In response, we propose a subset selection strategy that builds the training set upfront, without model input. This approach offers several practical benefits:
\begin{itemize}
    \item \textbf{Less experimental time.}  Because Perturb-seq assays are parallelizable, selecting all perturbations at once allows the experiments to conclude in approximately the time of a single active learning cycle—yielding months of real-world acceleration. 
    \item \textbf{Reduced operational complexity.} It avoids repeated cycles of coordination between the wet-lab and the model, model retraining, and acquisition computations, reducing both logistical overhead and potential errors. 
\item \textbf{Greater stability.} By decoupling the selection strategy from model initialization it ensures consistent perturbation sets across runs, improving reproducibility and reusability of collected data.
\end{itemize}

The only search space we can use to define subset selection is the knowledge graph, which is defined between genes based on prior knowledge and is used for message passing by the model.
To this end, we propose a graph-based method to build the train set for one of the state-of-the-art models, GEARS \citep{roohani2024predicting} as seen in in Fig.~\ref{fig:df}, although it can be combined with any graph-based predictor. 
We draw inspiration from recent findings on the relationship between train and test nodes to develop a method that selects the genes that maximize the model's reach on the graph.
We show that this selection increases the amount of genes trained, which in tandem increases the probability of having non-random embeddings in the receptive field of a test node. The criterion is proven submodular, and it is optimized greedily to build the training set with a near-optimal theoretical guarantee.
Our contributions are as follows:  
\begin{itemize}
\item A novel subset selection method, \textsc{GraphReach}, that selects genes such that the supervision signal to the embeddings is maximized. 
\item An empirical comparison including two active learning methods (one of which is state of the art in our problem), indicating that \textsc{GraphReach} achieves at least substantially faster training and increased stability without essentially sacrificing accuracy.
\end{itemize}

%% file: related.tex
\section{Related Work}
\label{sec:related}

The body of related work can be broadly divided into two categories. The first concerns optimal experimental design, where the objective is to choose a set of perturbations from a large action space in order to maximize an expected outcome variable, such as T-cell activation~\citep{mehrjou2021genedisco}. Methods such as Bayesian optimization and online learning have been applied in this context~\citep{pacchiano2022neural, lyle2023discobax,pacchiano2022neural}, and are evaluated based on the number of high-reward interventions discovered. These algorithms are not applicable in our case because they require a plethora of experiment rounds (e.g., 40), which are feasible with CRISPRi experiments but not when it is combined with PerturbSeq. 
Moreover, they focus on the experimental success of the retrieved selection rather then the efficacy of the final learning model, which typically predicts scalar values such as phenotypes not multidimensional vectors as in our problem.

The second branch of relative literature focuses on efficient training strategies for models that predict the gene expression profile following perturbation in single cells.
In this setting, the aim is not to optimize a causal phenotype but to train a model that generalizes well to unseen perturbations. This is particularly useful in Perturb-seq experiments, which measure the transcriptomic effects of perturbations via single-cell RNA sequencing. Due to the high cost and long duration of each experimental cycle, active learning methods have been proposed to efficiently select the most informative perturbations for training. 
The closest method to our approach is \textsc{IterPert}~\citep{huang2024sequential}, which uses active learning and prior multimodal knowledge to build a train set for GEARS. Since each iteration can take 3-5 weeks in the wet-lab, the number of iterations is diminished to 5 in contrast to ~50 in optimal experimental design \citep{mehrjou2021genedisco}. The problem is addressed from the perspective of active learning under budget~\citep{hacohen2022active} with the inclusion of prior imaging and perturb-seq studies. In fact, prior multimodal data was so effective, that it produced state of the art results without active learning, i.e., the model \textit{\textsc{IterPert-prior-only}}. However, such data is not commonly available. 

Our method differs from prior work in several ways. First, we propose a method that selects the perturbations prior to model training, thereby requiring only a single experimental round to get the labels. Since Perturb-seq experiments can be parallelized, this reduces the typical duration of training the model from 5 months to approximately one month. Second, unlike \textsc{IterPert}, our approach relies only on an open-access ontology, making it applicable across settings where curated multimodal data is unavailable. Third, by removing model-driven selection, our methods achieve stable and reproducible gene selection. This improves the reusability and interpretability of the resulting data. Finally, the proposed method achieves competitive accuracy compared to the state-of-the-art active learning method that does not use prior multimodal data, \textsc{TypiClust} \citep{huang2024sequential}.

\begin{figure*}[h!]
    \centering
    \includegraphics[width=0.75\linewidth]{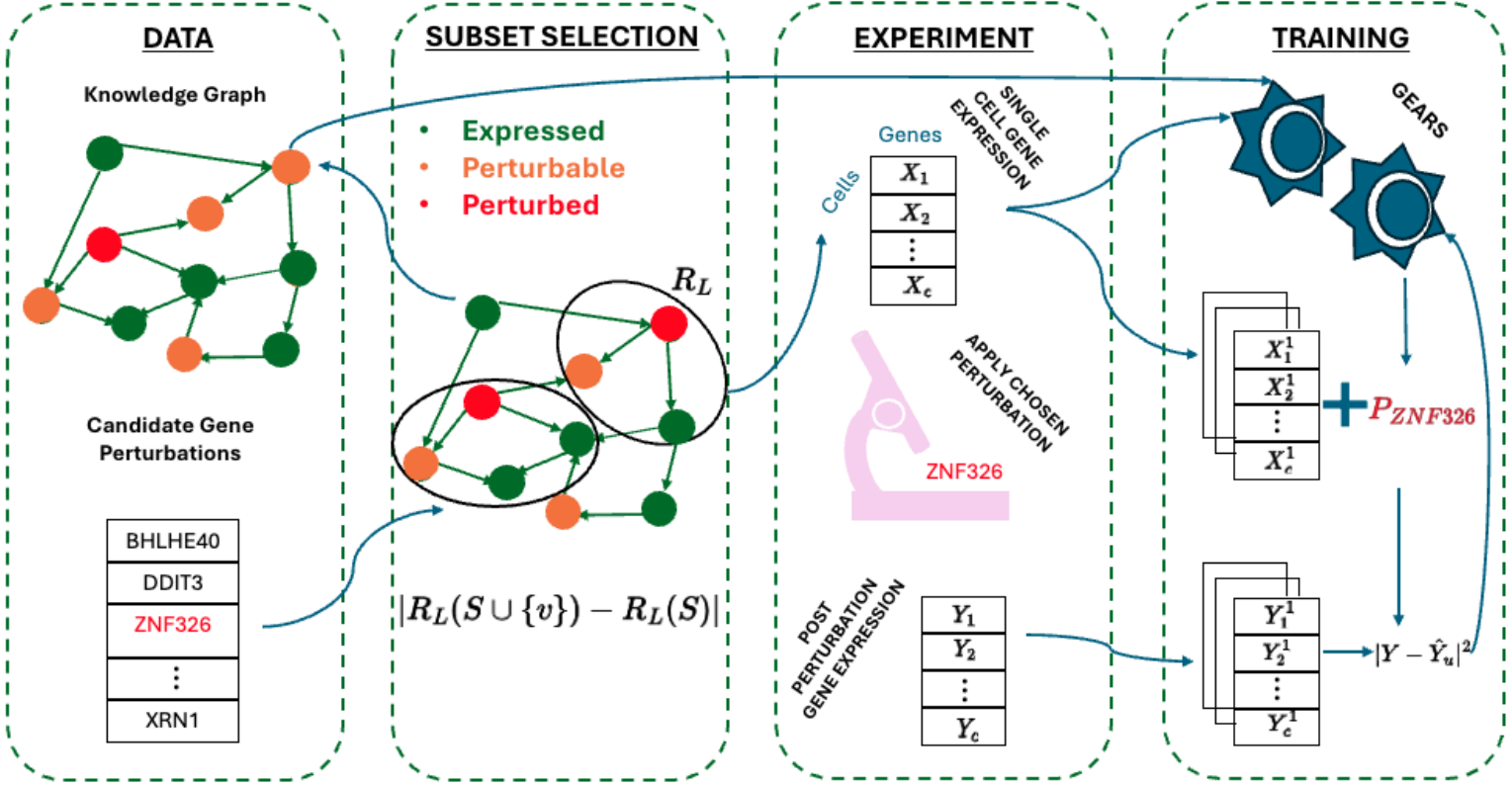}
    \caption{An overview of the subset selection methodology. Our sole input is the knowledge graph from GO51~\citep{roohani2024predicting} and a list of candidate perturbations. The subset selection algorithm, such as \textsc{GraphReach}, selects the set of gene perturbations, and they are given to the wet lab for the experimental part. Finally, the single-cell gene expression data is given to GEARS for training and validation.}
    \label{fig:flow}
\end{figure*}

%% file: methodology.tex
\section{Methodology}
\label{sec:method}
In this section, we formulate the problem, clarify the theoretical motivation for our approaches and introduce the method for subset selection.


\subsection{Formulation}
We are given a gene knowledge graph \( G = (V, E) \), where \( |V| = N \) and \( |E| = M \) with adjacency matrix $\mathbf{A}$, and each node represents a gene, some of which correspond to candidate gene perturbations. Our goal is to select a train set \( S \subset V \), under a budget constraint \( |S| = b \) with \( b \ll N \), to experimentally perturb in a wet lab setting.

For each selected gene \( u \in S \), the perturbation experiment yields paired measurements of pre- and post-perturbation gene expression: \( \mathbf{X} \in \mathbb{R}^{c \times g} \) for \( c \) single-cell profiles and \( g \) expressed genes (not to be confused with the perturbed genes), and corresponding post-perturbation outputs \( \mathbf{Y}_u \in \mathbb{R}^{c \times g} \). Thus, the perturbation effect is modeled as an additive change \( \hat{P}_u \in \mathbb{R}^g \) to $\mathbf{X}$ such that \( \mathbf{\hat{Y}}_u = \mathbf{X}  + 1_{c} \hat{P}_u^\top\) where \(1_{c} \in \mathbb{R}^{c \times 1}\) is a column vector of ones, and the model learns a mapping \( M: V \rightarrow \mathbb{R}^g \).
 Note that the perturbation is run on multiple cells with varying gene expressions.

The model parameters \( \theta^S \), trained only on the selected set \( S \), represent a graph neural network (GNN) such as GEARS (\cite{roohani2024predicting}). The loss function is the mean squared error between the predicted and the observed expression:

\begin{equation}
    \mathcal{L}(\mathbf{X}, \mathbf{Y}, u; \theta^S) = \| \mathbf{Y} - (\mathbf{X} + 1_{c} \hat{P}_u^\top) \|^2,
\end{equation}
where \( \hat{P}_u = M(u; \theta^S) \). The problem we address is to select the most informative subset of genes to perturb, such that the trained model generalizes well to unseen perturbations \( U \subseteq V \setminus S \). Formally, we aim to optimize:

\begin{equation}
    \underset{S \subset V,\ |S| = b}{\arg\min} \ \sum_{u \in U} \mathcal{L}( \mathbf{X},  \mathbf{Y}_u, u; \theta^S).
\end{equation}
\noindent


\subsection{Subset Selection to Maximize Supervision}

Since we refrain from utilizing the model's predictions, we turn to methodology that resembles traditional density-based active learning, such as Core-sets \citep{sener2017active}, which aims to cover as much of the input space as possible. 
Such methods contend that if we assume each training sample to cover all samples in a ball with radius $\delta$ around them
, then choosing the training set to maximize coverage of the dataset reduces the generalization error. We also know from the literature that labeling a node results in an information gain for every other node in its receptive field, i.e., $k$-hop neighborhood \citep{wang2020unifying,zhang2022information}. Finally, we know that the generalization is improved as we reduce the geodesic distance between the training and the test nodes \citep{ma2021subgroup}, which is performed implicitely when the train set maximizes its reach.
These findings support our intuition that covering a larger part of the graph with training nodes 
should be beneficial for the model.

In the case of GEARS, let $\mathbf{H}_\ell$ represent the gene embeddings from layer $\ell$ of the simplified graph convolution (SGC) used to predict the perturbation $\hat{P}$.
Assuming a single SGC layer for clarity (while omitting subsequent layers between $\mathbf{H}$ and the loss $\mathcal{L}$), we have: 
\begin{align}
    \mathbf{E} &= \mathbf{O} \mathbf{W}_0, \label{eqn:embedding}\\
    \hat{\mathbf{A}}&= \left(\mathbf{D}^{-1/2} (\mathbf{A} + \mathbf{I}d)\mathbf{D}^{-1/2}\right)^k,\label{eqn:GSO}\\
    \mathbf{H}&= \hat{\mathbf{A}} \mathbf{E} \mathbf{W}_1, \label{eqn:SGC}
\end{align}
 where $\mathbf{D}$ is the diagonal degree matrix, $\mathbf{W}_0 \in \mathbb{R}^{N \times d}$ is the embedding lookup table, $\hat{\mathbf{A}}$  is the normalized adjacency with self-loops to the power of $k=1$ (which is the default parameter chosen in the GEARS model), $\mathbf{W}_1\in \mathbb{R}^{d \times d'}$ are the SGC's weights, and $\mathbf{O} \in \mathbb{R}^{N\times N}$ is a row-wise one-hot encoding of the candidate gene perturbations.
 We base our methodology on the fact that message-passing GNNs update only the representations in $\mathbf{W}_0$ of nodes that are within the receptive field of the supervised nodes. To formalize this, we now state a proposition in which we explicity calculate the gradient with respect to a particular node representation. 

\begin{proposition} \label{prop:gradient}
    For the model defined in Equations \ref{eqn:embedding}, \ref{eqn:GSO} and \ref{eqn:SGC}, the gradient of the loss with respect to an individual gene embedding $\mathbf{W}_0[i]$ is defined as 
    \begin{align}
    \frac{\partial \mathcal{L}}{\partial \mathbf{W}_0[i]}
    &= \sum_{j=1}^N \hat{\mathbf{A}}[j,i] \cdot 
    \left( \frac{\partial \mathcal{L}}{\partial \mathbf{H}[j]} \, \mathbf{W}_1^\top \right).
\end{align}
Therefore, the gradient of $\mathbf{W}_0[i]$ depends only on the gradients of its neighbors in $\hat{\mathbf{A}}$.
\end{proposition}
The proof of Proposition \ref{prop:gradient} can be found the appendix. From Proposition \ref{prop:gradient} we can conclude that the supervision signal is propagated solely to nodes reachable from the train set, meaning that the identity embedding $\mathbf{W}_0$ that is learned for each gene, is updated only if it lies within the receptive field of supervised nodes.

This means that if we choose the train set such that we maximize the reachable nodes, we will maximize in tandem the number of embeddings $\mathbf{W}_0$ adjusted to the supervision signal. This is beneficial because, since GEARS is inherently semi-supervised, expanding the supervision allows for the representations of test samples or their neighbors to get updated. Having non-random embeddings in the receptive field of the test nodes should in principle lead to better generalization. It should be noted that while our analyzes relies on the architecture of GEARS, we expect the conclusion to generalize to a number of similar semi-supervised GNNs.
This clarifies how covering larger parts of the graph can potentially improve the prediction.
Therefore, to maximize the reach of the supervision signal, we should aim to maximize the number of nodes reached by the training set $S$.


\subsection{\textit{\textsc{GraphReach}} } 
We define our subset selection criterion as a function that maximizes the number of nodes reached by the node set $S$, i.e, the receptive field.
Specifically for a new node $v$, the criterion is defined as the additional reachability that $v$ adds in the set of train perturbations $S$ 
\begin{equation}
\label{eq:criterion}
    \alpha(v,S) = |R_L(S \cup \{v\}) - R_L(S)|,
\end{equation}
where $R_L$ is the set of nodes reachable from set $S$, based on the number of SGC layers $L$ in the definition of GEARS:
\begin{equation}
\label{eq:reach}
R_L(S) = \{ v \in V \mid \mathbf{A}^\ell_{uv} > 0 \text{ for } u \in S, 1 \leq \ell \leq L \}.
\end{equation}

The subscript $L$ is omitted in subsequent discussion as it remains stable throughout all experiments.
The criterion function selects greedily the node $v$ maximizing $\alpha(v, S)$ in every iteration. This is fundamentally suboptimal \citep{baykal2021low,kirsch2019batchbald} but like many active learning methods~\citep{lyle2023discobax,pacchiano2022neural,tigas2022interventions,sussex2021near}, we will prove our criterion is monotonic and submodular to achieve a guarantee that the result will be at least $1-1/e$ close to the optimal~\citep{nemhauser1978analysis}. 

\begin{proposition}
    The reachability function $R$ as defined in Eq.~\ref{eq:reach} is monotonic and submodular.
\end{proposition}

The reader is referred to the Appendix for the whole proof. 
Consider two sets $S_t \subseteq S_{t+1} \subseteq V$ and any node $v \notin S_{t+1}$. 
Since $S_t \subseteq S_{t+1}$, every node reachable from $S_t$ is also reachable from $S_{t+1}$, meaning $R(S_t) \subseteq R(S_{t+1}) \Rightarrow~|R(S_t)|\leq |R(S_{t+1})| $, which proves monotonicity. 
By definition,  adding a node earlier ($\alpha(v, S_t)$) adds at least as many reachable nodes as adding it later ($\alpha(v, S_{t+1})$), since the nodes in the cut of $S_t$ and $S_{t+1}$ can belong to $R(v)$ as well: $R(v) \cap (R(S_{t+1}) \setminus R(S_{t}))\geq 0$. This means that any node that appears in $R(v)$ and in $(R(S_{t+1}) \setminus R(S_{t}))$ was new for $R(S_{t})$ but is not new for $R(S_{t+1})$. Hence, the additional reachability that $v$ can bring to $S_{t+1}$ compared to the one it can bring to $S_{t}$ is diminished. This sketches out the proof that the reachability is submodular. 


Despite the theoretical guarantee, the greedy approach to maximizing reachability over the graph requires still requires testing every available perturbation for every addition during a cycle, i.e., $\mathcal{O}(NT)$ where $T$ is the total training budget in samples. 
This can be substantial depending on the graph or the experiment's scope. To this end, we employ a cost effective lazy forward approach to accelerate the acquisition step without loss of accuracy~\citep{leskovec2007cost}. The final subset selection algorithm, called  \textit{\textsc{GraphReach}}  from graph reachability, is shown in Alg.~\ref{alg:graphreach}, and an overview of the overall step-by-step subset selection approach can be seen in Fig.~\ref{fig:flow}. 
Finally, it should be noted that the method is easily extended to combinatorial perturbations. The set $S$ includes the candidate sets of perturbations, and the reachability $R(S)$ is the total number of unique nodes reached by the joint candidate node set. If two subsets of $S$ overlap, then the joint set ensures the individual gene appears only once.


\begin{algorithm}[t]
\caption{ {\sc GraphReach}}
\label{alg:graphreach}
\begin{algorithmic}[1]
\Input  budget $B$, graph $G$
\State Train set $S = \emptyset$
\State Criterion $\Delta[v] = \alpha(v,\emptyset)$ for all $v \in V$
\State sort($\Delta$)
\State Heap $\mathcal{L}$ with gene perturbations prioritized by $\Delta$
\While{$\mathcal{L} \neq \emptyset$ and $|S|\leq B$}
    \State $v$  = $\mathcal{L}[0]$
    \State $u$  = $\mathcal{L}[1]$
    \State $\delta =  \alpha(v,S)$ 
    \If{$\delta > \Delta[u]$}
        \State $S = S \cup \{v\}$
        \State $\mathcal{L} = \mathcal{L}[1:]$
    \Else
        \State $\Delta[v] = \delta$
        \State sort($\Delta$) and rearrange $\mathcal{L}$ accordingly
    \EndIf
\EndWhile
\Output $S$
\end{algorithmic}
\end{algorithm}


%% file: experiments.tex
\section{Experiments}
\label{sec:exp}
As mentioned in the introduction, subset-selection is significantly faster than active learning due to high-throughput genomics platforms like Perturb-seq being explicitly designed for parallelized experiments. This translates to a many-fold acceleration in our context. Besides speed, here we quantify the changes in other dimensions of the problem, addressing these questions:
\begin{itemize}
    \item \textbf{Accuracy}: How is the generalization of the model affected by the training procedure?
    \item \textbf{Stability}: How much do the proposed genomic experiments change throughout different runs?
    \item \textbf{Robustness}: Is noise in the knowledge graph able to erode the accuracy of the model?
\end{itemize}

To this end, in this section, we first describe the data used for the experiments, including the gene perturbation datasets and the knowledge graph, the benchmarks used for comparison, as well as the experimental design and the evaluation methods. Afterwards, we showcase the performance of the methods and analyze them to derive conclusions about the soundness of the proposed techniques. The code of the experiments, along with details on computing infrastructure, is provided in the supplementary material.

\subsection{Data}

We test our methods in four single-cell genomic datasets stemming from PerturbSeq experiments, following the literature \citep{roohani2024predicting} as seen in Tab. \ref{tab:datasets}. The datasets are diverse in terms of the number of perturbations and the number of samples per perturbation. The \textbf{Norman} dataset includes combinatorial perturbations (up to 2 genes) and the rest contain data on single perturbations. 

\begin{table}[t]
   \caption{The number of distinct gene perturbations in each datasets (single or combinatorial) along with the average number of cells (samples) per perturbation.}\label{tab:datasets}
   \begin{center}
    \begin{tabular}{lll}
         \textbf{Dataset} & \textbf{\makecell{\textsc{Perturbation}\\\textsc{Number}}  }&  \textbf{\makecell{\textsc{Average}\\\textsc{Cell Count}}} \\
         \hline
         \textbf{Adamson} & 81 & 800 \\
         \textbf{K562} & 1,087 & 150\\
         \textbf{RPE1} & 1,535 & 105 \\
         \textbf{Norman}  & 277 & 322
    \end{tabular}
\end{center}\end{table}

Akin to the literature, the graph is based on pathway information from GO51 \citep{gene2004gene}. Each gene is associated with a number of pathway GO51 terms. The Jaccard similarity between the sets of pathways of two genes, i.e., the fraction of shared pathways, is used to calculate the strength of the edge between them.
The graph is sparsified by keeping a predefined number of the most important neighbors for each node. The final graph contains over 9,800 nodes and 200,000 edges, exactly as in \citep{roohani2024predicting}. 
It should be noted that GEARS utilizes the whole knowledge graph to learn representations and uses the gene perturbation datasets for supervision only in a semi-supervised manner. \textsc{GraphReach} follows suit and utilizes the whole graph for selection. 

\subsection{Benchmarks}
As stated in Section \ref{sec:related}, there is currently still a lack of methods addressing our tackled problem. \textit{\textsc{IterPert}} does not operate without multi-modal data and hence cannot be utilized in our setting. Thus, we rely on these benchmarks:
\begin{itemize}
\item \textsc{Baseline} represents the random selection from the available gene perturbations. This is the prevalent practice because of its speed and simplicity. It represents the vanilla usage of GEARS.
    \item \textsc{ACS-FW} \citep{pinsler2019bayesian} is a Bayesian batch active learning model. It selects perturbations such that the new posterior distribution of the model's parameters approximates the expected posterior when the whole dataset is available. 
    We utilized the version from the BMDAL\-reg library \citep{holzmuller2023framework}, which was one of the top-performing methods in similar experiments \citep{huang2024sequential}.
    \item \textsc{TypiClust} \citep{hacohen2022active} is the state-of-the-art active learning method on this problem as it has outperformed 8 active learning models \citep{huang2024sequential}. 
    The algorithm clusters the candidate perturbations based on the final graph layer representation from GEARS. Within each cluster, the typicality is quantified as the inverse of the average distance between each sample and an example's $K$-nearest neighbors, with $K=20$. The most typical sample is selected. 
\end{itemize}

\begin{figure*}[h!]
    \centering
    \includegraphics[width=1\linewidth]{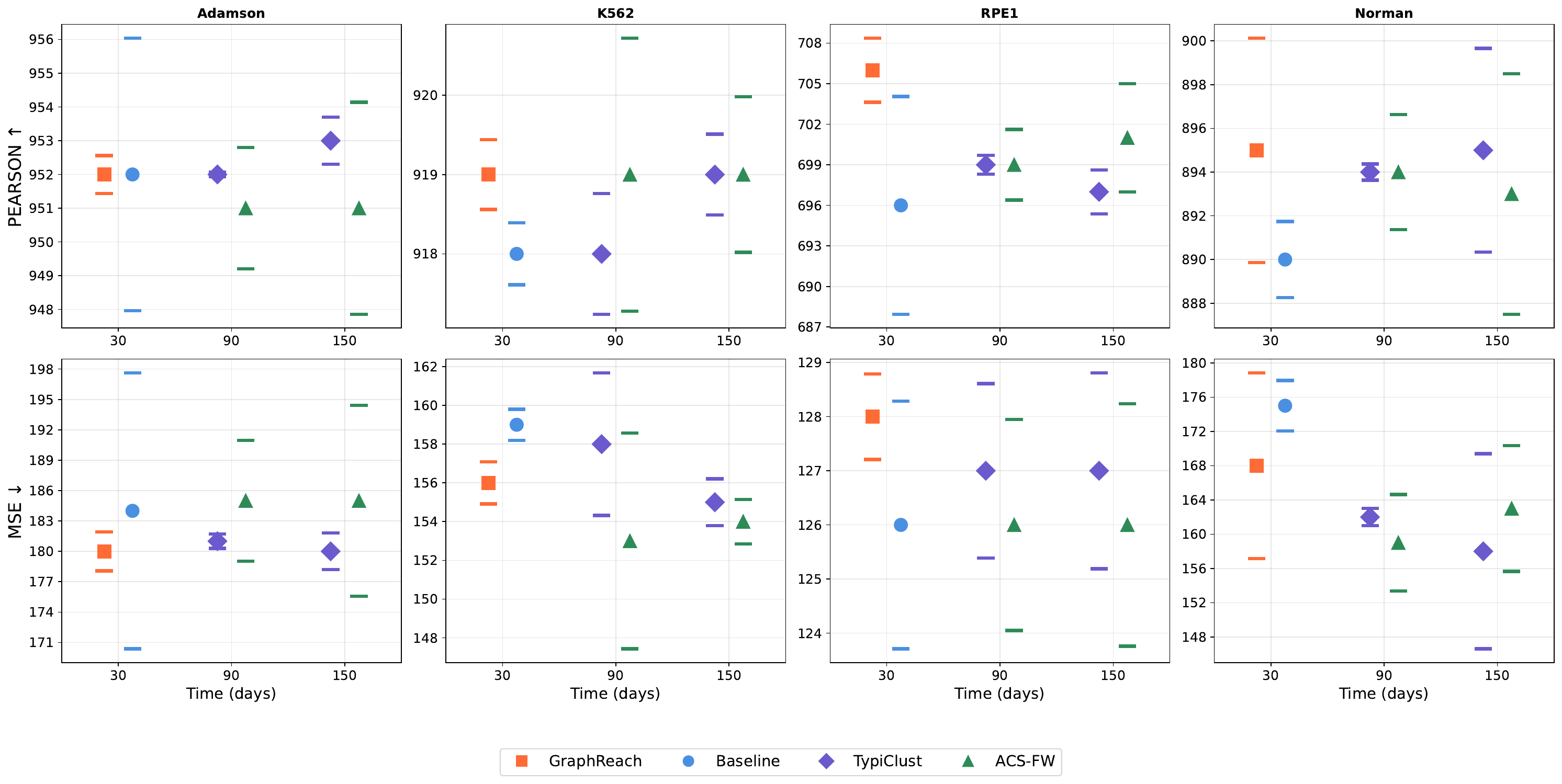}
    \caption{Accuracy in the test set with confidence intervals compared to the total physical time required for the training due to wet-lab iterations. \textsc{GraphReach} and \textsc{Baseline} do not require model-input hence they are trained with only one cycle of wet lab experiments taking 30 days. \textsc{TypiClust} and \textsc{ACS-FW} inherently need numerous cycles, and they are each run for 90 and 150 days to highlight the role that the number of cycles plays in the accuracy.} 
    \label{fig:time_performance}
\end{figure*}

\subsection{Design}

For \textsc{GraphReach} and \textsc{Baseline}, the train and validation sets are defined before the acquisition of the single-cell data and they require no interaction with the wet-lab. Thus, they are expected to take around 30 days, which corresponds to the time the PerturbSeq experiment takes, since running \textsc{GraphReach} at the beginning and training GEARS at the end takes negligible time. For all methods, 25\% of the gene perturbations are used for training and 5\% for validation, 10\% are kept for testing, leaving 60\% of the dataset unutilized. This is imperative to perform an effective comparison between the selected train sets. If we increase the train set size, the methods converge unavoidably to very similar final gene selections because the available choices are limited. Hence, we leave sufficient space between choices to highlight the differences between the strategies. 
We use a 10-fold cross validation to adjust the 10\% test set and take the average performance. The whole experiment is repeated for 3 different random seed initializations (including model initialization and data splits).

For the active learning methods, in accordance to ~\citep{huang2024sequential}, we keep the initial 10\% test set constant throughout the cycles to ensure there is no interaction between the strategies and the evaluation. 
We train the model a number of times sequentially, each time adding to the train and validation set a percentage of the available perturbations until they reach 25\% and 5\%, respectively. 
In order to quantify the relationship between the training cycles and final model accuracy, active learning methods are run for 3 and 5 cycles, the latter being the main choice in the literature. The models with 5 cycles receive 5\% of the perturbations in each cycle for training and 1\% for validation, while for 3 cycles the numbers are 8.3\% and 1.7\% respectively. Models trained on 5 cycles are expected to take 150 days, and 3 cycles correspond to 90 days. The methods are evaluated based on the performance of GEARS trained on the final cycle. It should be noted that no active learning method can go less than 60 days, as 1 training cycle (30 days) means no input from the model.

Following the literature, the evaluation metrics for performance are Pearson correlation and MSE, which is computed on the top 20 most differentially expressed genes, i.e., the genes that exhibit the biggest expression differences in the experiment. This is a common practice since the vast majority of the genes have zero expression, meaning the results would be skewed and the differences minuscule had we computed them for more genes \citep{roohani2024predicting}.
We evaluate stability by quantifying how much the chosen gene perturbations diverge between different seed initializations for each method. To this end, we quantify the consistency of each method by measuring the overlap between the resulting gene sets using the average pairwise Jaccard similarity defined in the appendix.
To evaluate robustness to noise, we performed the same experiments where \textsc{GraphReach} runs on corrupted versions of the knowledge graph. Specifically, we delete 5\% and 10\% of the edges and add the same amount of fake ones, to represent a common setting of errors in the graph.  
In all experiments, we use the default GEARS parameters and implementation\footnote{\url{https://github.com/snap-stanford/GEARS/tree/master}}.

\subsection{Results}




\subsubsection{Accuracy}

The overall tradeoff between accuracy and efficiency is visualized in Fig. \ref{fig:time_performance}. 
For Pearson correlation, we observe that \textsc{GraphReach} achieves the strongest performance on average, as it performs on par or better than the benchmarks in three of the four datasets, despite taking a fraction of the time. 
We see that it outperforms the \textsc{Baseline}, the only benchmark with similar efficiency, in 7 out of 8 evaluations, and that the latter is significantly more uncertain in terms of confidence intervals.

On average, the active learning methods perform on par with each other, with \textsc{TypiClust} being better in \textbf{Adamson} and \textbf{Norman} and \textsc{ACS-FW} being better in the rest. 
Another observation for \textsc{ACS-FW} is that it remains constant for 3 and 5 cycles for \textbf{Adamson}, 3 cycles are better than 5 in \textbf{Norman} while the results are split in \textbf{K562} and \textbf{RPE1}. We believe that this happens because training on 3-cycles increases the amount of samples the model sees per cycle and this allows GEARS to provide better uncertainty estimates which are integral for \textsc{ACS-FW}. In contrast, \textsc{TypiClust} tends to deteriorate as we constrain the days with only one exception in \textbf{RPE1} for Pearson correlation.
That said, \textsc{ACS-FW} has considerably larger confidence intervals (competing with the \textsc{Baseline}) throughout all experiments, possibly due to the sensitivity to GEARS' uncertainty estimation. We analyze further this instability in the coming sections. We thus deem \textsc{TypiClust} the second best method, as it is considerably more stable and achieves competitive performance.

Note that, as mentioned above, active learning requires by default a number of cycles to run and hence it can not achieve similar efficiency to \textsc{GraphReach} and we can not compare them head-to-head. However, if we constrain the experiments to up to 90 days, \textsc{GraphReach} performs better than active learning in 5 out of 8 evaluations. Given that on average active learning deteriorates as we constrain the days, we can deduce that  \textsc{GraphReach} showcases the best tradeoff between accuracy and efficiency.

\subsubsection{Stability}
The average Jaccard similarity between the perturbation sets selected through all cycles can be found in Tab. \ref{tab:avg_jaccard}.
The random seed affects the model and the k-folds of the data, hence the set of genes retrieved by \textsc{GraphReach} differ solely due to different splits. In contrast, active learning methodologies exhibit variability across different runs that approximate the \textsc{Baseline}, which is a random selection. 
This instability implies that the resulting gene sets are tightly coupled to model-specific biases, limiting their reusability for downstream analyses or integration with other studies, and indicating that the selection process is driven more by model bias rather than by a signal in the data.

\begin{table}[t]

    \caption{Average Jaccard similarity on gene selections through different random seed initializations.}
    \label{tab:avg_jaccard}
    \centering
    \setlength{\tabcolsep}{1mm}
    \begin{tabular}{lrrrr}
        \textbf{Data} & \textbf{\makecell{\textsc{Base}\\\textsc{line}}}& \textbf{\makecell{\textsc{ACS}\\\textsc{FW}}} & \textbf{\makecell{\textsc{Typi}\\\textsc{Clust}} }& \textbf{\makecell{\textsc{Graph}\\\textsc{Reach}}} \\
        \hline
        \textbf{Adamson} & 0.15 & 0.15 & 0.15 &  \textbf{0.75} \\
        \textbf{K562} & 0.10 & 0.13 & 0.10  & \textbf{0.78} \\
        \textbf{RPE1} & 0.10 & 0.13 & 0.10 & \textbf{0.76}  \\
        \textbf{Norman} & 0.11 & 0.13 & 0.15  &  \textbf{0.95}\\
    \end{tabular}
\end{table}

\begin{table}[h!]
\caption{Performance under different noise levels in the knowledge graph, as represented by random removal and addition of edges. Results are rounded up to the nearest integer.}
\centering
\label{tab:noise}
\begin{tabular}{lcrr}
\textbf{Dataset} & \textbf{Noise (\%)} & \textbf{Pearson} (↑) & \textbf{MSE} (↓) \\
\hline
\multirow{3}{*}{\textbf{Adamson}} 
 &0 & $952 \pm 1$ & $180 \pm 2$ \\
 & 5& $951 \pm 1$ & $181 \pm 6$ \\
 & 10& $947 \pm 1$ & $196 \pm 8$ \\
\hline
\multirow{3}{*}{\textbf{K562}} 
 &0& $919 \pm 1$ & $156 \pm 1$ \\
 & 5& $919 \pm 2$ & $154 \pm 1$ \\
 & 10& $920 \pm 2$ & $153 \pm 2$ \\
\hline
\multirow{3}{*}{\textbf{RPE1}} 
 &0& $706 \pm 3$ & $128 \pm 1$ \\
 & 5& $701 \pm 4$ & $129 \pm 1$ \\
 & 10& $704 \pm 6$ & $126 \pm 2$ \\
\hline
\multirow{3}{*}{\textbf{Norman}} 
 &0& $895 \pm 5$ & $168 \pm 10$ \\
 & 5& $893 \pm 13$ & $170 \pm 18$ \\
 & 10& $894 \pm 13$ & $163 \pm 13$ \\
\end{tabular}
\end{table}

\subsubsection{Robustness}

 The results from the robustness experiments are in Tab. \ref{tab:noise}. We observe that while \textsc{GraphReach} underperforms in noisy settings in the \textbf{Adamson} dataset, it actually stays close to or even surpasses the original in the rest. We hypothesize that this happens because random edges may have a relatively high probability of connecting previously distant parts of the graph and may therefore further increase the coverage that \textsc{GraphReach} is able to achieve. So, we observe satisfactory empirical robustness of \textsc{GraphReach}. 
 Similar effects of random perturbations on approximation quality have been observed in studies of vertex cover and related graph problems \cite{shi2018runtime}.

%% file: appendix.tex
\appendix
\section*{Appendix}
\label{sec:appendix}


\section{Proof of Proposition 1}

We begin by considering the gradient of the loss $\mathcal{L}$ with respect to the identity embeddings $\mathbf{W}_0$ as a function of $\frac{\partial \mathcal{L}}{\partial \mathbf{H}}$:
\begin{align}
         \frac{\partial \mathcal{L}}{\partial \mathbf{W}_0} &= \mathbf{O}^\top \left( \frac{\partial \mathcal{L}}{\partial \mathbf{E}} \right) ,\\
      \frac{\partial \mathcal{L}}{\partial \mathbf{E}}&= \hat{\mathbf{A}}^\top \,  \left(  \frac{\partial \mathcal{L}}{\partial \mathbf{H}} \, \mathbf{W}_1^\top \right) .
\end{align}
Thus, the full gradient becomes:
\begin{align}
   \frac{\partial \mathcal{L}}{\partial \mathbf{W}_0} 
    = \mathbf{O}^\top \, \hat{\mathbf{A}}^\top \, \frac{\partial \mathcal{L}}{\partial \mathbf{H}} \, \mathbf{W}_1^\top.
\end{align}
For an individual embedding row $\mathbf{W}_0[i]$ and since $\mathbf{O}$ is one-hot, this expands to:
\begin{align}
    \frac{\partial \mathcal{L}}{\partial \mathbf{W}_0[i]}
    &= \sum_{j=1}^N \hat{\mathbf{A}}[j,i] \cdot 
    \left( \frac{\partial \mathcal{L}}{\partial \mathbf{H}[j]} \, \mathbf{W}_1^\top \right),
\end{align}
which shows that the gradient of $\mathbf{W}_0[i]$ depends only on the gradients of its neighbors in $\hat{\mathbf{A}}$. However, $\frac{\partial \mathcal{L}}{\partial \mathbf{H}[j]}=0$ if $j\notin S$. 

\section{Proof of Proposition 2}

Following the definitions of the main paper, let $R$ be the set function of reachability in the graph, and we have subsets of nodes $S_t \subseteq S_{t+1}$.
We can start from the definition of submodularity:
\begin{equation} \label{eqn:StartingPoint}
R(S_t\cup \{u\}) \setminus R(S_t) \supseteq R(S_{t+1} \cup \{u\}) \setminus R(S_{t+1}) 
\end{equation}
and continue by reformulating the right and left-hand side of Eq. \ref{eqn:StartingPoint} as follows,
\begin{align*}
R(S_t\cup \{u\}) \setminus R(S_t) 
&= \big(R(S_t) \cup R(\{u\})\big) \setminus R(S_t) \\
&= R(\{u\}) \setminus R(S_t), \\
R(S_{t+1} \cup \{u\}) \setminus R(S_{t+1}) 
&= \big(R(S_{t+1}) \cup R(\{u\})\big) \setminus R(S_{t+1}) \\
&= R(\{u\}) \setminus R(S_{t+1}).
\end{align*}
Plugging these reformulations back into Eq. \ref{eqn:StartingPoint} produces,
\begin{equation}\label{Eq:ReformulatedSubmod}
R(\{u\}) \setminus R(S_t) \supseteq R(\{u\}) \setminus R(S_{t+1}).
\end{equation}

We shall now derive the reformulated definition of submodularity in Eq. \ref{Eq:ReformulatedSubmod} to show that reachability is submodular.

We start with the fact that by definition we have \(S_t\subseteq S_{t+1}\) and graph reachability \(R\) is monotone, consequently:
\begin{equation}
\quad R(S_t) \subseteq R(S_{t+1}).
\label{eq:rseq}
\end{equation}

Now recall the anti-monotonicity property of set difference (as illustrated below for sets $X, Y$, and $Z$):
\begin{equation}
X \subseteq Y \quad \Longrightarrow \quad Z \setminus X \supseteq Z \setminus Y.
\end{equation}
So if we add a set difference on Eq. \ref{eq:rseq} with $R(\{u\})$, we get the desired:
\begin{equation}
R(\{u\}) \setminus R(S_t) \supseteq R(\{u\}) \setminus R(S_{t+1}).
\end{equation}

Thus,
\begin{equation}
R(S_t\cup \{u\}) \setminus R(S_t) 
\supseteq
R(S_{t+1} \cup \{u\}) \setminus R(S_{t+1}),
\end{equation}

which proves that reachability is submodular.

\section{Computational Time}
The computational time is negligible in our setting, because each sample selection takes less than a minute. That said, \textit{\textsc{GraphReach}} is around 500 times faster than \textit{\textsc{TypiClust}}, meaning it scales significantly better with the number of available perturbations. This is important for experiments without predefined candidate perturbations, where the search space can increase from hundreds to tens of thousands.

\section{Infrastructure} 
The computing infrastructure used for the experiments reported in this paper includes a 13th Gen Intel(R) Core(TM) i9-13900K CPU with 24 cores, an NVIDIA GeForce RTX 4070 with 24GB and a CUDA Version: 12.2, and a RAM of 32 GB on an Ubuntu 22.04.